\documentclass[12pt]{article}
\usepackage{amssymb,amsmath,amsthm}
\usepackage{graphicx}
\usepackage{float}
\numberwithin{equation}{section}

\def\<{\langle}
\def\r>{\rangle}

\def\Hc{{\cal H}}

\def\Pc{{\cal P}}
\def\Sc{{\cal S}}

\begin{document}
\title{Complex Energies and Beginnings of Time Suggest a Theory of
  Scattering and Decay}
\author{A.~Bohm\thanks{bohm@physics.utexas.edu}\\
  Department of Physics\\
  The University of Texas at Austin\\
  Austin, TX 78712 \and
  P.~Kielanowski\thanks{kiel@physics.utexas.edu}\\
  Departamento de F\'{\i}sica, CINVESTAV, Mexico\\
  \and
  S.~Wickramasekara\thanks{wick@rice.edu}\\
  Department of Physics and Astronomy, Rice University\\
  Houston, TX 77005}

\maketitle
\begin{abstract}
  Many useful concepts for a quantum theory of scattering and decay
  (like Lippmann-Schwinger kets, purely outgoing boundary conditions,
  exponentially decaying Gamow vectors, causality) are not well
  defined in the mathematical frame set by the conventional (Hilbert
  space) axioms of quantum mechanics. Using the Lippmann-Schwinger
  equations as the takeoff point and aiming for a theory that unites
  resonances and decay, we conjecture a new axiom for quantum
  mechanics that distinguishes mathematically between prepared states
  and detected observables. Suggested by the two signs $\pm i\epsilon$
  of the Lippmann-Schwinger equations, this axiom replaces the one
  Hilbert space of conventional quantum mechanics by two Hardy spaces.
  The new Hardy space theory automatically provides Gamow kets with
  exponential time evolution derived from the complex poles of the
  $S$-matrix. It solves the causality problem since it results in a
  semigroup evolution. But this semigroup brings into quantum physics
  a new concept of the semigroup time $t=0$, a beginning of time. Its
  interpretation and observations are discussed in the last section.
\end{abstract}

\section{Introduction}\label{sec1}
Quantum theory falls into, roughly, two categories~\cite{bohm1}:
\begin{enumerate}
\item[{I.}] The description of spectra and structures of micro-physical
  systems
\item [{II.}]Scattering and decay phenomena
\end{enumerate}

The distinction between the two categories is primarily one between
two ways one looks at the physical objects, rather than a separation
of physics into two different areas. The first is used for stable
states and also for slowly decaying states when the finiteness of
their lifetime is ignored. The second is used for rapidly decaying
states and resonance phenomena. The notions of slow and fast are not
defined by a time scale in nature but by the capabilities of the
experimental apparatuses that we choose or are forced to use in a
particular experiment. For instant, the singly excited states of atoms
and molecules are mostly treated like stable states whereas the doubly
excited states (Auger states) are mostly treated as resonances or
decaying states. However, when one does the calculations of the
energies of the Auger states (e.g., of He) one ignores that they
decay~\cite{bohm1}.

The same holds in nuclear physics and in high energy physics. When one
is interested only in the spectra and the structure of relativistic
particles, one ignores their lifetimes even though the different
states of the same multiplet can have lifetimes that are orders of
magnitudes apart. (E.g., one can measure the lifetime of $\Omega^-$
but one cannot measure the lifetime of $\Delta$~\cite{frauenfelder}.
The existence and properties of $\Delta$ are determined from lineshape
measurements and lifetime was chosen as the inverse of the lineshape
width on the basis of some theoretical ideas/approximations for which
a theory did not exist \cite{levy}.)

For category I (spectra and structure), one uses a theory of
stationary states and time symmetric (reversible) evolutions. The
energy values are discrete and the time evolution is unitary and the
superpositions are effectively finite. Such systems are well described
by conventional quantum mechanics in the Hilbert space $\Hc$. Infinite
superpositions are handled by perturbative methods (of discrete
spectra).

The second category (scattering and decay) deals with continuous
energy spectra and predominantly asymmetric time evolutions. If one
wants to use energy eigenstates, the continuous energy values already
require more than what the conventional axioms of quantum mechanics
are able to accommodate. This has been overcome by introducing the
Dirac kets $\left|E\r>\right.$, which -if they are mathematically
defined at all- are defined as functionals on the Schwartz space. With
this definition, energy wave functions $\psi(E)=\<E|\psi\r>$ do not
constitute the entire Hilbert space of (Lebesgue) square-integrable
functions, but only the subspace of infinitely differentiable, rapidly
decreasing functions, i.e., Schwartz space functions.

The introduction of Dirac kets augments the conventional axiomatic
framework of quantum mechanics based on the Hilbert space and leads to
the Gelfand triplet $\Phi\subset\Hc\subset\Phi^\times$, where $\Phi$
is the Schwartz space and
$\left|E\r>\right.\in~\Phi^\times$~\cite{bohm1}. However, the Gelfand
triplet based on the Schwartz space is not sufficient to obtain a
theory that includes scattering and decay. The reason is that the
dynamical (Schr\"odinger or Heisenberg) equations, when defined as
differential equations in the Schwartz space of wavefunctions,
integrate to a continuous group of evolution operators, much like the
unitary group solution of these equations in the Hilbert space.

In contrast, resonances and decaying states have been intuitively
associated to an asymmetric ``irreversible'' time
evolution~\cite{merzbacher}. Thus, they require a time asymmetric
theory, and in the absence of such a mathematical theory, their
description can only be approximate and must contain contradictions.
If one is guided by the Hilbert space mathematics, one always runs
into problems with a quantum theory of resonances and decay; in
particular, Gamow vectors with exponential decay do not exist in the
Hilbert space. Therefore, in the heuristic treatment of scattering
theory, one just ignored the mathematical subtleties of the Hilbert
space. In particular, one worked with mathematically undefined kets
$\left|E^\pm\r>\right.$, used an infinitesimal imaginary energy part
$\pm i\epsilon$ to obtain, respectively, the incoming and outgoing
solutions of the Lippmann-Schwinger equations~\cite{lippmann}, and
distinguished between ``states at time $t'<t_0=$ time defined by
preparation'' and ``states characteristic of the experiment'',
observed at $t''>t_0$~\cite{feynman}. One restricted by fiat the time
$t$ in $e^{iHt}$ to $t\geq0$~\cite{gell-mann}, and for decaying
states, one postulated purely outgoing boundary
conditions~\cite{peierls}, undisturbed by the fact that it was in
conflict with the unitary group evolution $-\infty<t<\infty$, a direct
consequence of the conventional Hilbert space axioms of quantum
mechanics (by the Stone-von Neumann theorem \cite{stone, neumann}).
These heuristic methods were successful for physical applications, but
when one compared them with the mathematical consequences of the
Hilbert space axiom, one had contradictions. Examples of these are:
the exponential catastrophe in which Gamow vectors and unitary time
evolution were mutually contradictory~\cite{bohm2} and references
therein; deviations from the exponential decay law~\cite{bohm3}; and
problems with (Einstein) causality~\cite{fermi}.

It is thus clear that one has to go beyond the mathematical theory
which has worked for Category I problems. But many of the empirical
notions, like Gamow states and Lippmann-Schwinger kets, have been very
successful for the descriptions of scattering and decay, and their
successful features need to be preserved when they are incorporated
into the new rigorous theory. However, other mathematical consequences
of the conventional axioms need to be eliminated.  This means we
require a new hypothesis which preserve the successful features and
alter the conflicting fallouts from the conventional theory. New
mathematical entities will have to be defined, which we will call
again by their old names, like Lippmann-Schwinger kets or Gamow kets,
but they will now have new features and are constituents of a
consistent theory of resonance scattering and decay. The new
mathematical hypothesis will be conjectured taking the useful features
of these heuristic notions as the starting point.

\section{Conventional Quantum Theory Conflicts\\
  with the Lippmann-Schwinger Equations}\label{sec2}
         
By conventional quantum theory, we hereon mean not only the usual
axioms~\cite{neumann} in terms of the Hilbert space mathematics, but
also the Dirac formalism mathematically justified by, as stated above
in the Introduction, a Gelfand triplet of the Schwartz space. The
axiomatic framework of conventional quantum mechanics consists of the
following:
\begin{itemize}
\item[{(A1)}] One distinguishes (physically) between observables
  represented by self-adjoint operators (e.g., $A$, $\Lambda$
  (positive operators), or vectors $\psi$ if
  $\Lambda=|\psi\r>\<\psi|$) and states represented by trace
  class operators (e.g., $W$ or vectors $\phi$ if $W=|\phi\r>\<\phi|$).\\
  The quantities compared with experimental data are the Born
  probabilities $\Pc_{W(t)}(\Lambda)={\rm Tr}(W(t)\Lambda)={\rm
    Tr}(W\Lambda(t))$, or, in the special case $W=|\phi\r>\<\phi|$ and
  $\Lambda=|\psi\r>\<\psi|$, $\Pc_{\phi(t)}(\psi)=
  \left|\<\psi|\phi(t)\r>\right|^2=\left|\<\psi(t)|\phi\r>\right|^2$.
  That is,
  \begin{equation}
    \Pc_{\phi(t)}(\psi)=\left|\<\psi|\phi(t)\r>\right|^2=
    \left|\<\psi(t)|\phi\r>\right|^2\simeq\frac{N_1(t)}{N}\nonumber
  \end{equation}
  The experimental quantities $\frac{N_1(t)}{N}$ are the ratios of
  large integers (detector counts which necessarily change in time in
  discrete steps). On the other hand, every mathematical theory is an
  idealization and thus quantum theory also idealizes to continuous
  time translations, in consequence of which the calculated Born
  probabilities ${\cal P}_{\phi(t)}(\psi)$ change continuously in time
  in a particular way.  The equality between the two quantities ${\cal
    P}_{\phi(t)}(\psi)$ and $\frac{N_1(t)}{N}$ is approximate --and
  the sign $\simeq$ expresses this aspect of the statistical character
  of quantum mechanical predictions-- and the meaning of the
  continuity for $\phi(t)$ or ${\cal P}_{\phi(t)}(\psi)$ as a function
  of $t$ is a mathematical choice.
\end{itemize}
In conventional quantum mechanics one makes this choice by identifying
\begin{itemize}
\item[{(A2)}] The set of states $\{\phi\}$= The set of observables
  $\{\psi\}=\Hc=$ Hilbert space
\end{itemize}
In Dirac's formalism one assumes in addition that
\begin{itemize}
\item[{(A3)}] for every observable, e.g., $H$, one has a complete set
  of eigenkets $|E\r>$ such that
  \begin{itemize}
  \item [{(3a)}]$H|E\r>=E|E\r>$
    \noindent and
  \item[{(3b)}] Every vector, state $\phi$ or observable $\psi$, is a
    continuous superposition of the eigenkets extending over all
    ``physical values'' $0\leq E<\infty$:
    \begin{equation}
      \phi=\sum_{j,j_3,\eta}\int
      dE|E,j,j_3,\eta\r>\<E,j,j_3,\eta|\phi\r>\nonumber
    \end{equation}
    (here $j,j_3$ and $\eta$ are some additional quantum numbers
    representing the degeneracy of the eigenkets with energy $E$.)
  \end{itemize}
\end{itemize}

Nearly everyone discussing the foundations of quantum mechanics
\cite{foundations} distinguishes between states and observables as
asserted by (A1) above. The Hilbert space axiom (A2) is already in
conflict with this hypothesis (A1) because the content of (A1) is a
basic distinction between a state and an observable. Also, the
hypothesis (A3), the Dirac formalism, is not possible within the
framework of the Hilbert space axiom (A2) since neither (3a) nor (3b)
is well defined as a vector identity in the Hilbert space when $E$ is
a continuous parameter.

One can overcome this difficulty and make (A3) mathematically tenable
by restricting the vectors $\{\phi\}$ and $\{\psi\}$ to a subspace
$\Phi$ of the Hilbert space and constructing a Gelfand triplet,
$\{\phi\}=\{\psi\}=\Phi\subset\Hc\subset\Phi^\times$. With this choice
of $\Phi$, the eigenkets $|E\r>$ can be defined as the elements of the
dual space $\Phi^\times$ and (3b) can be proved as the nuclear
spectral theorem. As stated above, if the Schwartz space is chosen for
$\Phi$ so that energy wavefunctions
$\phi(E)=\<E|\phi\r>=\overline{\<\phi|E\r>}$ are smooth and rapidly
decreasing at infinity, then the dual space $\Phi^\times$, which
consists of continuous anti-linear functionals on $\Phi$, is realized
by the space of tempered distributions. Therefore, in this
representation, the eigenkets $|E\r>$ find realization as tempered
distributions.

In scattering theory, one has in-states $\{\phi^+\}$ and
out-observables $\{\psi^-\}$ (which are usually called out-states). An
in-state $\phi^+$ is prepared at $t\rightarrow-\infty$ in the
asymptotic region as the interaction-free in-states $\phi^{\rm in}$
such that
\begin{equation}
  \phi^{\rm in}\rightarrow\phi^+\nonumber\\
\end{equation}
Similarly, for $t\rightarrow\infty$, the out-observable $\psi^-$
becomes the interaction free out-observable $\psi^{\rm out}$ which
describes a measurable property in the asymptotic region:
\begin{equation}
  \psi^-\rightarrow\psi^{\rm out}\nonumber
\end{equation}

The superscripts $\pm$ of state vectors $\phi^+$ and $\psi^-$ have
their origins in the labels of the eigenkets $|E^\pm\r>$ of the full
Hamiltonian $H=H_0+V$,
\begin{equation}
  H|E^\pm\r>=E|E^\pm\r>\label{2.1}
\end{equation}  
The Dirac basis vector expansion of (3b) above holds for every
$\phi^+$ and every $\psi^-$ in terms of the eigenkets $|E^+\r>$ and
$|E^-\r>$, respectively:
\begin{subequations}
  \label{2.2}
  \begin{equation}
    \tag{\ref{2.2}+}
    \{\phi^+\}\ni\phi^+=\sum_{jj_3\eta}\int_0^\infty
    dE|Ejj_3\eta^+\r>\<^+Ejj_3\eta|\phi^+\r>\label{2.2+}
  \end{equation}
  \begin{equation}
    \tag{\ref{2.2}$-$}
    \{\psi^-\}\ni\psi^-=\sum_{jj_3\eta}\int_0^\infty
    dE|Ejj_3\eta^-\r>\<^-Ejj_3\eta|\psi^-\r>\label{2.2-}
  \end{equation}
\end{subequations}
The eigenkets $|E^\pm\r>$ of the full Hamiltonian in \eqref{2.1} are
also assumed to be the plane-wave solutions to the Lippmann-Schwinger
equations
\begin{subequations}
  \label{2.3}
  \begin{equation}
    |E^\pm\r>=|E\r>+\lim_{\epsilon\rightarrow0}\frac{1}{E-H_0\pm
      i\epsilon}V|E^\pm\r>=\Omega^\pm|E\r>
    \tag{\ref{2.3}$\pm$}
\end{equation}
\end{subequations}
where $|E\r>$ fulfill the eigenvalue equation $H_0|E\r>=E|E\r>$ for
the ``free Hamiltonian'' $H_0$ of \eqref{2.1}.

As seen from \eqref{2.1}, the eigenkets $|E^+\r>$ and $|E^-\r>$ both
correspond to the same eigenvalue $E$, but (\ref{2.3}$\pm$) shows that
they fulfill different boundary conditions expressed by $+i0$ and
$-i0$.

In scattering theory, the set of functions that are admitted to serve
as energy wave functions in (\ref{2.2}$\pm$),
\begin{subequations}
  \label{2.4}
  \begin{equation}
    \tag{\ref{2.4}+}
    \phi^+(E)=\<Ejj_3\eta^+|\phi^+\r>=\<E|\phi^{\rm in}\r>\label{2.4+}
  \end{equation}
  and
  \begin{equation}
    \tag{\ref{2.4}$-$}
    \psi^-(E)=\<Ejj_3\eta^-|\psi^-\r>=\<E|\psi^{\rm out}\r>\label{2.4-}
  \end{equation}
\end{subequations}
are usually assumed to be the same set of smooth functions as the
functions $\<Ejj_3\eta|\phi\r>$ that appear in the basis vector
expansion hypothesis (A3b). That is,
\begin{subequations}
  \label{2.6}
  \begin{equation}
    \{\phi^+(E)\}=\{\psi^-(E)\}=\{\phi(E)\}=\text{Schwartz function
      space}\label{2.6a}
  \end{equation}
  For the vectors, this means
  \begin{equation}
    \{\phi^+\}=\{\psi^-\}=\Phi\subset\Hc\subset\Phi^\times\label{2.6b}
  \end{equation}
\end{subequations}
(where $\Phi$ is dense in $\cal H$).  The assumption
$\{\phi^+\}=\{\psi^-\}=\Phi$ (or, the version ${\Hc}^{\rm
  in}={\Hc}^{\rm out}=\Hc$) is known in scattering theory and quantum
field theory as the assumption of asymptotic completeness.

The time evolution of the state $\phi^+(t)$ is given by the
Schr\"odinger equation
\begin{subequations}
  \label{2.7}
  \begin{equation}
    \tag{\ref{2.7}+}
    i\hbar\frac{d\phi^+(t)}{dt}=H\phi^+(t)\label{2.7+}  
  \end{equation}
\end{subequations}
The solution to this equation under the Hilbert space boundary
condition of assumption (A2) above is
\begin{subequations}
  \label{2.8}
  \begin{equation}
    \tag{\ref{2.8}+}
    \phi^+(t)=e^{-iHt}\phi^+,\ \text{with}\ -\infty<t<\infty\label{2.8+}
  \end{equation}
\end{subequations}
The time evolution of the observable
$\Lambda(t)=\left|\psi^-(t)\r>\<\psi^-(t)\right|$ is given by the
Heisenberg equation of dynamical motion
\begin{equation}
  \tag{\ref{2.7}$-$}
  \frac{d\Lambda(t)}{dt}=\frac{-i}{\hbar}[\Lambda(t),H],\quad\text{or
    by}\quad i\hbar\frac{d\psi^-(t)}{dt}=-H\psi^-(t)\label{2.7-}
\end{equation}
The solution of this equation under the Hilbert space boundary
condition of assumption (A2) is
\begin{equation}
  \label{2.8-}
  \tag{\ref{2.8}$-$}
  \Lambda(t)=e^{iHt}\Lambda e^{-iHt},\quad\text{or}\quad
  \psi^-(t)=e^{iHt}\psi^-\quad
  \text{with}\; -\infty<t<\infty
\end{equation}
If $\{\phi^+\}$ and $\{\psi^-\}$ are assumed to be a Hilbert space and
if the Hamiltonian $H$ is a self-adjoint operator, then, by the
well-known Stone-von Neumann theorem \cite{stone}, \eqref{2.8+} and
\eqref{2.8-} are necessarily the unique solutions to the dynamical
equations in the Schr\"odinger and Heisenberg pictures, \eqref{2.7+}
and \eqref{2.7-}.  Moreover, this theorem asserts that the operators
$e^{-iHt}$ and $e^{iHt}$ are unitary for each $-\infty<t<\infty$ and
that the mappings $t\rightarrow e^{-iHt}\phi^+$ and $t\rightarrow
e^{iHt}\psi^-$ are continuous. It is noteworthy that Stone's theorem
requires the (norm complete) Hilbert space
$\{\phi^+\}=\{\psi^-\}=\Hc$, in contrast to, say, \eqref{2.6} above.
However, it is possible to show that the solutions (\ref{2.8}$\pm$)
hold for all $-\infty<t<\infty$ also for the Schwartz space completion
of \eqref{2.6}, although there are subtle mathematical differences
between the two cases (A2) and \eqref{2.6}~\cite{wick}.
  
If the solutions (\ref{2.8}$\pm$) hold for the vectors $\phi^+$ and
$\psi^-$, then it follows, by duality, that the eigenkets $|E^+\r>$
and $|E^-\r>$ behave much like $\psi^-$ and $\phi^+$, respectively.
That is,
\begin{subequations}
  \label{2.9}
  \begin{equation}
    \<\phi(t)|E^+\r>=\<e^{-iHt}\phi^+|E^+\r>=\<\phi^+|e^{iH^\times t}|E^+\r>
    =e^{iEt}\<\phi^+|E^+\r>\label{2.9a}
  \end{equation}
  Or, as an eigenvalue equation between functionals,
  \begin{equation}
    e^{iH^\times t}|E^+\r>=e^{iEt}|E^+\r>,\quad -\infty<t<\infty\label{2.9b}
  \end{equation}
\end{subequations}
Likewise,
\begin{subequations}
  \label{2.10}
  \begin{equation}
    \<\psi^-(t)|E^-\r>=\<e^{iHt}\psi^-|E^-\r>=\<\psi^-|e^{-iH^\times t}|E^-\r>
    =e^{-iEt}\<\psi^-|E^-\r>\label{2.10a}
  \end{equation}
  Or, as an eigenvalue equation between functionals,
  \begin{equation}
    e^{-iH^\times t}|E^-\r>=e^{-iEt}|E^-\r>,\quad -\infty<t<\infty\label{2.10b}
  \end{equation}
\end{subequations}
In \eqref{2.10a} and \eqref{2.10b}, $H^\times$ is the uniquely defined
extension of $\bar{H}=H^\dagger$ to the space $\Phi^\times$.  It is
clear that \eqref{2.9b} and \eqref{2.10b} depend on the time evolution
of $\phi^+$ and $\psi^-$, given by \eqref{2.8+} and \eqref{2.8-}. The
latter equations depend on the assumption that $\phi^+$ and $\psi^-$
are elements of the Schwartz space $\Phi$ of \eqref{2.6}. Therefore,
if \eqref{2.9b} and \eqref{2.10b} hold, then $|E^\pm\r>$ must be
Schwartz space kets, i.e., functionals on the Schwartz space, meaning
that $\phi^+(E)=\<^+E|\phi^+\r>$ and $\psi^-(E)=\<^-E|\psi^-\r>$ are
infinitely differentiable, rapidly decreasing functions on the {\em
  real} (and positive) energy axis.

This requirement on $|E^\pm\r>$, however, is in contradiction with the
requirement that $|E^\pm\r>$ be solutions of the Lippmann-Schwinger
equations (\ref{2.3}$\pm$) which contain the complex energies $E\pm
i\epsilon$. As already mentioned, there is a physical distinction
between the vectors $\phi^+$ and $\psi^-$ as being related to
experimentally accessible $\phi^{\rm in}$ and $\psi^{\rm out}$ for
$t\rightarrow-\infty$ and $t\rightarrow\infty$, respectively. As we
shall see in the next section, these {\em asymmetric} boundary
conditions in time are what give rise to the limits
$\epsilon\rightarrow0^+$ and $\epsilon\rightarrow0^-$ in
(\ref{2.3}$\pm$) that define the $\pm$ signs in the kets $|E^\pm\r>$.

\section{What the Lippmann-Schwinger Equations 
Suggest}\label{sec3}

It is the term $\pm i\epsilon$ in (\ref{2.3}$\pm$) which tells us that
the Lippmann-Schwinger kets
$|E^\pm\r>=\lim_{\epsilon\rightarrow0}|E\pm i\epsilon\r>$ cannot be
ordinary Dirac kets (Schwartz space functionals). The infinitesimals
$\pm i\epsilon$ indicate that the energy wave functions
$\<\phi^+|E^+\r>$ and $\<\psi^-|E^-\r>$ must not only be Schwartz
space functions of the real variable $E$, as asserted by the axiom
\eqref{2.6}, but they must also be limits of functions defined on some
region of the upper and lower complex plane of $E$. It is simplest to
assume that $\<\phi^+|E^+\r>$ and $\<\psi^-|E^-\r>$ are boundary
values of {\em analytic} functions defined on such a region in the
(open) upper complex half-plane $\mathbb{C}_+$ and lower complex
half-plane $\mathbb{C}_-$, respectively. As the complex semi-plane in
energy, one takes the second (or higher) Riemann surface of the
analytic $S$-matrix. Thus, we have the following basic hypothesis
which replaces \eqref{2.6}:
\begin{subequations}
  \label{3.1}
  \begin{equation}
    \tag{\ref{3.1}+}
    \text{Functions}\ \phi^+(E)=\<^+E|\phi^+\r>=
    \overline{\<\phi^+|E^+\r>}\ \text{have analytic
      extensions into}\ {\mathbb{C}_-}\label{3.1+}
  \end{equation}
  and
  \begin{equation}
    \tag{\ref{3.1}$-$}
    \text{Functions}\ \psi^-(E)=\<^-E|\psi^-\r>=
    \overline{\<\psi^-|E^-\r>}\ \text{have analytic
      extensions into}\ {\mathbb{C}_+}\label{3.1-}
  \end{equation}
\end{subequations}

To make (\ref{2.3}$\pm$) possible, the analytic extensions of
\eqref{3.1+} and \eqref{3.1-} must exist at least on a small strip
below and above on the real energy axis (i.e., the physical scattering
energies). We shall generalize this to the hypothesis that the
analytic extensions of the energy wave functions should exist on the
entire upper and lower energy half-planes.

The requirement (\ref{3.1}$\pm$) is not inconsistent with the Schwartz
space hypothesis of \eqref{2.6}. Rather, (\ref{3.1}$\pm$) strengthens
\eqref{2.6}. However, the stronger condition (\ref{3.1}$\pm$) is not
consistent with the solutions (\ref{2.8}$\pm$) of the dynamical
equations (\ref{2.7}$\pm$), obtained as consequences of the weaker
condition \eqref{2.6}. Likewise, the time evolutions equations
\eqref{2.9b} and \eqref{2.10b}, which one universally assumes for
(all) energy eigenkets, also do not hold under the hypothesis
(\ref{3.1}$\pm$).

As stated above, the requirements of \eqref{3.1} are supplementary to
the usual hypothesis of quantum mechanics. Thus, the wave functions
$\phi^+(E)$ and $\psi^-(E)$ are still assumed to be, for instant,
smooth, rapidly decreasing and square integrable.  The simultaneous
requirements of analyticity and square integrability introduces
certain (unexpected) restrictions into the theory. For instant, it can
be shown~\cite{bohm4,gadella} that these requirements can be met for
the time translated functions \eqref{2.9a} and \eqref{2.10a} only if
$t\geq0$.\footnote{Actually, this feature of time evolution can be
  seen from a simple heuristic argument that goes as follows. If the
  time translated function $\<\phi^+(t)|E^+\r>$, just like the
  function $\<\phi^+|E^+\r>$ is the square integrable boundary value
  function of an analytic function defined in the upper half-plane,
  then for $E=E+i\epsilon$, we have
  $\<\phi^+(t)|E+i\epsilon\r>=e^{i(E+i\epsilon)t}\<\phi^+|E+i\epsilon\r>$.
  Since $\epsilon$ is positive,
  $e^{i(E+i\epsilon)t}\<\phi^+|E+i\epsilon\r>$ is bounded for
  arbitrary values of $\epsilon$ only if $t$ is positive. A similar
  argument holds for the time translation of the observable wave
  functions $\<\psi^-(t)|E^-\r>$ of \eqref{2.10a}. The rigorous proof
  is given in text following \eqref{3.7.5}.} Since the time
translation equations \eqref{2.9b} and \eqref{2.10b} are derived from
\eqref{2.9a} and \eqref{2.10a}, the conclusion $t\geq0$ also holds for
the kets $e^{\pm iEt}|E^\pm\r>$.

Thus, the first conclusion that we draw from the Lippmann-Schwinger
equations (\ref{2.3}$\pm$) is that the time evolution of the vectors
$\phi^+$ and $\psi^-$ in (\ref{2.2}$\pm$) should not be given by the
unitary group solution of the the dynamical equations
(\ref{2.7}$\pm$), but by the {\em semigroup} solution:
\begin{subequations}
  \label{3.2}
  \begin{equation}
    \tag{\ref{3.2}+}
    \phi^+(t)=e^{-iHt}\phi^+\quad\text{for}\ 0\leq t<\infty\
    \text{only}.\label{3.2+}
  \end{equation}
  \begin{equation}
    \tag{\ref{3.2}$-$}
    \psi^-(t)=e^{iHt}\psi^-\quad\text{for}\ 0\leq t<\infty\
    \text{only}.\label{3.2-}
  \end{equation}
\end{subequations}
From this we see that as a consequence of the $\pm i\epsilon$ in the
Lippmann-Schwinger equations (\ref{2.3}$\pm$), the $\{\phi^+\}$ and
$\{\psi^-\}$ given by the Dirac basis vector expansion
(\ref{2.2}$\pm$) are in general different mathematical quantities with
different (``conjugate'') semigroups (\ref{3.2}$\pm$) of time
evolution. The unitary group evolution (\ref{2.8}$\pm$) which follows
from \eqref{2.6} is in conflict with the Lippmann-Schwinger equations.
Time evolutions which are not in conflict with the Lippmann-Schwinger
equations (\ref{2.3}$\pm$) are (\ref{3.2}$\pm$).

Thus, on the basis of (\ref{3.1}$\pm$), we identify two different
vector spaces $\{\phi^+\}\not=\{\psi^-\}$, one for the states and the
other for the observables. The operators $e^{-iHt}$ and $H$ in
\eqref{3.2+} are operators defined in the vector space $\{\phi^+\}$.
Likewise, operators $e^{iHt}$ and $H$ in \eqref{3.2-} are operators
defined in the vector space $\{\psi^-\}$\footnote{To be precise in
  notation, one should distinguish between $H=H_-$, the restriction of
  the Hilbert space operator $\bar{H}$ to $\Phi_-=\{\phi^+\}$ and
  $H=H_+$, the restriction of the Hilbert space operator $\bar{H}$ to
  $\Phi_+=\{\psi^-\}$. For the sake of notational simplicity we will
  avoid this distinction whenever it does not lead to
  misunderstanding.}.  Now, from \eqref{3.1+} we know that the wave
functions $\phi^+(E)=\<^+E|\phi^+\r>$ corresponding to the vectors
$\phi^+$ are analytic in ${\mathbb{C}}_-$. Therefore, we call the
vector space $\{\phi^+\}\equiv\Phi_-$. Similarly, from \eqref{3.1-},
the wave functions $\psi^-(E)=\<\psi^-|E^-\r>$ are analytic in
${\mathbb{C}}_+$, and for this reason we call the vector space
$\{\psi^-\}\equiv\Phi_+$.  The two vector spaces $\Phi_\pm$ are then
two different subspaces of the Hilbert space $\Hc$ (and also of the
Schwartz space $\Phi$):
\begin{subequations}
  \label{3.3}
  \begin{equation}
    \tag{\ref{3.3}+}
    \phi^+\in\Phi_-\subset\Hc\label{3.3+}
  \end{equation}
  \begin{equation}
    \tag{\ref{3.3}$-$}
    \psi^-\in\Phi_+\subset\Hc\label{3.3-}
  \end{equation}
\end{subequations}

What remains now is to put additional conditions on the analytic
functions (\ref{3.1}$\pm$) such that the spaces $\Phi_\pm$ become
nuclear spaces. Then, the triplet of spaces
\begin{subequations}
  \label{3.4}
  \begin{equation}
    \tag{\ref{3.4}+}
    \Phi_-\subset\Hc\subset\Phi_-^\times\label{3.4+}
  \end{equation}
  \begin{equation}
    \tag{\ref{3.4}$-$}
    \Phi_+\subset\Hc\subset\Phi_+^\times\label{3.4-}
  \end{equation}
\end{subequations}
become Gelfand triplets, also known as Rigged Hilbert Spaces.  The
ordinary Dirac kets require one RHS \eqref{2.6b}. However, if the kets
are also to fulfill the Lippmann-Schwinger equations (\ref{2.3}$\pm$),
one needs the pair of RHS's, (\ref{3.4}$\pm$).  The $\Phi^\times_\pm$
in (\ref{3.4}$\pm$) are the dual spaces, consisting of continuous
anti-linear functionals on $\Phi_\pm$. The new kets $|E^\pm\r>$ have
then a well defined meaning as elements of the dual spaces
$\Phi_\pm^\times$, and the nuclear property of (\ref{3.4}$\pm$) allows
Dirac's basis vector expansion (\ref{2.2}$\pm$) to be established as
the nuclear spectral theorem of Gelfand {\em at al} and
Maurin~\cite{gelfand}.  The pair of Gelfand triplets \eqref{3.4} have
been constructed by Gadella~\cite{gadella} by choosing for the spaces
of wave functions \eqref{3.1} particular subspaces of Hardy
functions~\cite{duren} \footnote{ This choice is the following:
  \begin{subequations}
    \label{3.5}
    \begin{equation}
      \tag{\ref{3.5}+}
      \phi^+(E)\in\left.\Hc_-^2\cap\Sc\right|_{\mathbb{R}_+}\label{3.5+}
    \end{equation}
    \begin{equation}
      \tag{\ref{3.5}$-$}
      \psi^-(E)\in\left.\Hc_+^2\cap\Sc\right|_{\mathbb{R}_+}\label{3.5-}
    \end{equation}
  \end{subequations}
  Here, ${\Hc}^2_\pm$ denote Hardy class functions. $\Sc$ stands for
  the Schwartz space, and the symbol $\left.\right|_{{\mathbb{R}_+}}$
  represents the restriction of the domains of functions in
  $\Hc^2_\pm\cap\Sc$ to the positive real line, ${\mathbb{R}}_+$,
  assumed to be the range of scattering energy values. Loosely
  speaking, Hardy class functions $f^\pm\in\Hc_\pm^2$ are functions
  defined on the real line fulfilling the following two
  properties~\cite{bohm4,gadella,duren}:
  \begin{enumerate}
  \item $f^\pm(x)$ are point-wise limits of analytic functions
    $F^\pm(z)$ on ${{\mathbb{C}}}_\pm$, i.e.,
    $f^\pm(x)=\lim_{y\rightarrow0}F^\pm(x\pm iy)$
  \item The $f^\pm$ are square integrable,
    $\int_{-\infty}^\infty\left|f^\pm(x)\right|^2dx<\infty$
  \end{enumerate}
  The intersections $\Hc^2_\pm\cap\Sc$ ensure that the functions
  $\phi^+(E)$ and $\psi^-(E)$, in addition to having the desired
  analyticity properties for complex energies, are, for real energy
  values, infinitely differentiable and rapidly decreasing at
  infinity. Equally importantly, when defined as in \ref{3.5}, the
  nuclearity of the Schwartz space $\Sc$ can be used to define a
  topology for $\Phi_\pm$ so that these spaces are nuclear.  The
  one-to-one association of smooth Hardy functions for the energy wave
  functions in (\ref{3.5}$\pm$) is more restrictive than the
  analyticity of the wave functions in the small strip above or below
  the real axis, the weakest condition demanded by the
  Lippmann-Schwinger equations \eqref{2.3}. It is a mathematical
  idealization, like the idealization to Lebesgue square integrable
  functions in Hilbert space quantum mechanics. The Hardy space
  idealization, a refinement of the Hilbert space idealization, is
  better suited for quantum physics because it provides a mathematical
  distinction between states $\phi^+\in\Phi_-$ and observables
  $\psi^-\in\Phi_+$. It also provides a mathematical basis for the
  Lippmann-Schwinger integral equations, which incorporate the
  in-coming and out-going boundary conditions.}

Associated with an operator defined in the Hilbert space $\Hc$, there
exist two triplets of operators corresponding to the two triplets of
spaces in \eqref{3.5}. For instant, for the Hamiltonian $H$,
\begin{subequations}
  \label{3.6}
  \begin{equation}
    \tag{\ref{3.6}+}
    H_-\subset\bar{H}=H^\dagger\subset H_-^\times\label{3.6+}
  \end{equation}
  \begin{equation}
    \tag{\ref{3.6}$-$}
    H_+\subset\bar{H}=H^\dagger\subset H_+^\times\label{3.6-}
  \end{equation}
\end{subequations}
where $H_\mp$ are the uniquely defined restrictions
$\left.\bar{H}\right|_{{\Phi_\mp}}$ of the self-adjoint Hamiltonian
$\bar{H}$ to the dense subspace $\Phi_\mp$ of $\Hc$. The operators
$H_\mp^\times$ are the conjugate operators of $H_\mp$, which are
uniquely defined extensions of $H^\dagger$ to $\Phi_\mp^\times$. When
their meaning is clear from the context, we usually omit the
subscripts $\mp$ and superscript $\times$ in these various operators
and denote all of them simply by $H$.

Defining the Lippmann-Schwinger kets now as functionals on $\Phi_\mp$,
the $|Ejj_3\eta^\pm\r>\in\Phi^\times_\mp$ have analytic extensions
into the whole complex semi-plane ${\mathbb{C}}_\pm$ of the second
sheet of the $S$-matrix.
This property has turned out be to be very
important for the unified theory of resonances and decay.

In sum, we have conjectured the new hypothesis which
distinguishes mathematically between states and observables:\\

\begin{subequations}
  \label{3.7} Set of prepared states defined by preparation apparatus
    (accelerator), e.g., in-states
  \begin{equation}
    \tag{\ref{3.7}+}
    \{\phi^+\}=\Phi_-\subset\Hc\subset\Phi_-^\times\label{3.7+}
  \end{equation}

  Set of registered observables defined by registration apparatus
  (detector), e.g., out-states
  \begin{equation}
    \tag{\ref{3.7}$-$}
    \{\psi^-\}=\Phi_+\subset\Hc\subset\Phi_+^\times\label{3.7-}
  \end{equation}
\end{subequations}
We take \eqref{3.7} as a fundamental axiom which replaces the Hilbert
space axiom (A2) of Section \ref{sec2}.

The spaces $\Phi_\pm$ are two different dense subspaces of the same
Hilbert space $\Hc$. As stated above, the spaces $\Phi_\pm$ can be
understood as the abstract vector spaces whose realizations in terms
of energy wave functions have the smooth Hardy space property
(\ref{3.5}$\pm$). In other words, the space $\Phi_-$ is given by the
set of vectors $\{\phi^+\}$ whose Dirac vector expansion is given by
\eqref{2.2+}, where the ``coordinates $\<^+Ejj_3\eta|\phi^+\r>$ with
the continuous label'' $E$ (the analogue of the label $i=1,2,3$ in the
basis vector expansion $\vec{x}=\sum_{i=1}^3\vec{e}_ix^i$) are the
smooth Hardy functions $\phi^+(E)=\<^+Ejj_3\eta|\phi^+\r>$ with the
property \eqref{3.5+}. Similarly, the space $\Phi_+$ is the set of
vectors $\{\psi^-\}$ whose ``coordinates'' with respect to the
continuous basis $|Ejj_3\eta^-\r>$ are the smooth Hardy functions
\eqref{2.4-} with the property \eqref{3.5-}.  An immediate
mathematical consequence of the Hardy space axiom (\ref{3.5}$\pm$) is
that the solutions of the dynamical equations (\ref{2.7}$\pm$) have
the important (semigroup) property \eqref{3.2}:
\begin{subequations}
  \label{3.7.5}
  \begin{equation}
    \tag{\ref{3.7.5}+}
    \text{For $\phi^+(t)$ fulfilling Schr\"odinger's Eq.,}\ 
    \phi^+(t)=e^{-iH_-t}\phi^+\ \text{for}\ t\geq0\label{3.7.5+}
  \end{equation}
  \begin{equation}
    \tag{\ref{3.7.5}$-$}
    \text{For $\psi^-(t)$ fulfilling Heisenberg's Eq.,}\ 
    \psi^-(t)=e^{iH_+t}\psi^-\ \text{for}\ t\geq0\label{3.7.5-}
  \end{equation}
\end{subequations}

This semigroup time evolution (\ref{3.7.5}$\pm$) is a consequence of a
theorem of Paley and Wiener \cite{paley} (See also the appendix of
\cite{bohm5}) for Hardy class functions. The theorem states that if
$G_-(E)$ is a Hardy class function, then its Fourier transform
\begin{subequations}
  \label{paley1}
  \begin{equation}
    \tag{\ref{paley1}a}
    {\check{G}}_-(\tau)=\frac{1}{{\sqrt{2\pi}}}\int_{-\infty}^\infty dE
    e^{iE\tau}G_-(E)
    \label{paley1a}
  \end{equation}
  must fulfill the condition
  \begin{equation}
    \tag{\ref{paley1}b}
    {\check{G}}_-(\tau)=0\quad \text{for}\ -\infty<\tau<0
    \label{paley1b}
  \end{equation}
\end{subequations}
It further follows from the theorem that for any positive value of
$\tau$, say $|\tau_0|$, there exists a Hardy function
$G^{\tau_0}_-(E)\in{\cal{H}}\cap{\cal{S}}$ such that
\begin{equation} {\check{G}}^{\tau_0}_-(\tau)\not=0\quad \text{for}\
  0<\tau<|\tau_0|\label{paley1.1}
\end{equation}

Now, consider the Hardy space function $\<^+E|\phi^+\r>$ and the Hardy
space function $\<^+E|\phi^+(t)\r>$ of the time translated state
$\phi^+(t)$. Since $\phi^+(t)$ fulfills the Schr\"odinger equation
\eqref{2.7+}, $\phi^+(t)=e^{-iHt}\phi^+$ and its expansion
coefficients $e^{-iEt}\<^+E|\phi^+\r>$ in the basis vector expansion
\begin{eqnarray}
  \phi^+(t)=\int dE|E^+\r>\<^+E|\phi^+(t)\r>&=&\int
  dE|E^+\r>\<^+E|e^{-iHt}\phi^+\r>\nonumber\\
  &=&\int dE|E^+\r>\left(e^{-iEt}\<^+E|\phi^+\r>\right)\nonumber
\end{eqnarray}
as well as the expansion coefficient $\<^+E|\phi^+\r>$ in \eqref{2.2+}
must, according to \eqref{3.7+}, be a Hardy function of the lower
half-plane ${\mathbb{C}}_-$ if both $\phi^+$ and $\phi^+(t)$ are to
represent prepared states.  That is,
\begin{subequations}
  \label{paley2}
  \begin{equation}
    \tag{\ref{paley2}a}
    G_-(E)\equiv\<^+E|\phi^+\r>\in{\cal{H}}_-^2\cap{\cal{S}}\label{paley2a} 
  \end{equation}
  as well as
  \begin{equation}
    \tag{\ref{paley2}b}  
    G^t_-(E)\equiv
    e^{-iEt}\<^+E|\phi^+\r>\in{\cal{H}}_-^2\cap{\cal{S}} \label{paley2b}
  \end{equation}
\end{subequations}
It is an elementary property that the Fourier transform
${\check{G}}^t_-$ of the function \eqref{paley2b} is related to the
Fourier transform ${\check{G}}_-$ of the function \eqref{paley2a}:
\begin{equation} {\check{G}}^t_-(\tau)\equiv\frac{1}{\sqrt{2\pi}}\int
  dE e^{iE\tau}G_-^t(E)= \frac{1}{\sqrt{2\pi}}\int dE
  e^{iE(\tau-t)}G_-(E)=\check{G}_-(\tau-t)
\label{paley2.5}
\end{equation}
Now, if we want both $G_-(E)$ and $G^t_-(E)$ to be Hardy space
functions as in \eqref{paley2a} and \eqref{paley2b}, then it follows
from the Paley-Wiener theorem \eqref{paley1} that
\begin{subequations}
  \label{paley3}
  \begin{equation}
    \tag{\ref{paley3}a}
    \check{G}_-(\tau)=0\quad \text{for}\ -\infty<\tau<0
    \label{paley3a}
  \end{equation}
  and
  \begin{equation}
    \tag{\ref{paley3}b}
    {\check{G}}^t_-(\tau)=0\quad \text{for}\ -\infty<\tau<0
    \label{paley3b}
  \end{equation}
  But, becuase of \eqref{paley2.5}, we also have
  \begin{equation}
    \tag{\ref{paley3}c}
    {\check{G}}_-(\tau-t)=0\quad\text{for}\
    -\infty<\tau-t<0\label{paley3c}
  \end{equation}
\end{subequations}
From \eqref{paley3a} and \eqref{paley3c}, we have the simultaneous
conditions $-\infty<\tau<0$ and $-\infty<\tau<t$.  These two
requirements on $\tau$ are clearly satisfied for positive values of
$t$. If $t$ is negative, say $t=-|t|$, then the property
${\check{G}}^t(\tau)=\check{G}_-(\tau-t)=0$ is ensured only for
$-\infty<\tau<-|t|$, not for $-\infty<\tau<0$ as required by
\eqref{paley3b}.  In fact, from \eqref{paley1.1}, we see that there is
at least one function in the space ${\cal{H}}_-^2\cap{\cal{S}}$ for
which the condition \eqref{paley3b} is not fulfilled for
$-|t|<\tau<0$.  Therefore, $t\geq0$ must hold, and the time evolution
for the states $\phi^+(t)$ can only be defined for the semigroup
\eqref{3.7.5+}. A similar argument using the Hardy functions
${\cal{H}}_+^2\cap{\cal{S}}$ leads to the conclusion \eqref{3.7.5-}.

The conjugate operators\footnote{Note that the operators acting on the
  spaces $\Phi_\mp$ are labeled by the $\mp$ signs, e.g., $U_\mp,\
  H_\mp$. The $\mp$ signs labeling the spaces follow from the
  mathematicians' convention for the lower and upper Hardy class. The
  signs that label the vectors, on the other hand, follow from most
  physicists' notation of scattering theory and are opposite to those
  that label the spaces: $\phi^+\in\Phi_-,\ \psi^-\in\Phi_+,\
  |E^\pm\r>\in\Phi_\mp^\times$.}  of $U_\pm(t)$, defined by the
identities $\<U_-\phi^+|F^+\r>=\<\phi^+|U_-^\times F^+\r>$ for every
$\phi^+\in\Phi_-,\ F^+\in\Phi_-^\times$ and
$\<U_+\psi^-|F^-\r>=\<\psi^-|U_+^\times F^-\r>$ for every
$\psi^-\in\Phi_-,\ F^-\in\Phi_+^\times$, give the time evolutions in
the dual spaces $\Phi_\pm^\times$:
\begin{subequations} 
  \label{3.8}
  \begin{equation}
    \tag{\ref{3.8}+}
    U_-^\times(t)|F^+\r>=e^{iH_-^\times t}|F^+\r>,\ t\geq0,\
    F^+\in\Phi_-^\times\label{3.8+}
  \end{equation}
  \begin{equation}
    \tag{\ref{3.8}$-$}
    U_+^\times(t)|F^-\r>=e^{-iH_+^\times t}|F^-\r>,\ t\geq0,\
    F^-\in\Phi_+^\times\label{3.8-}
  \end{equation}
\end{subequations}
For the special case $F=|Ejj_3\eta^-\r>$ where
$H_+^\times|Ejj_3\eta^-\r>=E|Ejj_3\eta^-\r>$,
\begin{equation}
  U^\times_+(t)|Ejj_3\eta^-\r>=e^{-iH_+^\times t}|Ejj_3\eta^-\r>
  =e^{-iEt}|Ejj_3\eta^-\r>\ \text{for $t\geq0$ only.}\label{3.9}
\end{equation}

The set of operators $\{U_-(t)=e^{-iH_-t}:\ 0\leq t<\infty\}$ do not
form a group because there is no inverse operator
$\left(U_-(t)\right)^{-1}$ for every element of this set as required
by the group axioms. In contrast, for the set of unitary operators
$\{U(t)=e^{-iHt}:\ -\infty<t<\infty\}$ in the Hilbert space $\Hc$
there is an inverse operator $\left(U(t)\right)^{-1}=U(-t)$ for every
$U(t)$ so that the set constitutes a group. Aside from the absence of
inverse operators, the set of operators $\{U_-(t)=e^{-iH_-t}:\ 0\leq
t<\infty\}$ fulfills all other defining axioms of a group, and is
called a semigroup. Therefore, there are two different representations
of the time translation semigroup $0\leq t<\infty$ given by the
operators $U_\mp(t)=e^{\mp iH_\mp t}$ of (\ref{3.2}$\pm$) in the two
spaces $\Phi_\mp$. Likewise, the conjugate operators defined above in
(\ref{3.8}$\pm$) also furnish two representations of the time
translation semigroup $0\leq t<\infty$ in the dual spaces
$\Phi_\pm^\times$. In both of these cases, we have the condition
$t\geq0$ (because of the difference in sign on the right hand side of
the dynamical equations (\ref{2.7}$\pm$)).

The semigroup time evolution is an important consequence of the axiom
(\ref{3.7}$\pm$). This axiom makes it possible for the Hamiltonians
$H_\pm^\times$ to have eigenkets with complex eigenvalues. The
semigroup character of time evolution makes the probability densities
for complex energy eigenstates finite.  If one would force the unitary
time evolution (\ref{2.8}$\pm$) on these eigenstates with complex
energy, one would obtain infinite probabilities, which is the
well-known ``exponential catastrophe'' for the original Gamow wave
functions~\cite{bohm2}.
 
Under the new axiom (\ref{3.7}$\pm$), the Gamow state vector is
derived from the $S$-matrix pole at complex energy value
$z_R=E_R-i\Gamma/2$ as an eigenket (functional) $|z_Rjj_3\eta^-\r>\in
\Phi_+$ with generalized eigenvalue $z_R$~\cite{bohm3.5, bohm4,
  bohm5}.  In the construction of these Gamow kets, the eigenvalue
$z_R$ is the complex position of the $S$-matrix pole. Under the new
axiom (\ref{3.7}$\pm$), eigenkets of essentially self-adjoint
Hamiltonians with complex energy are now well defined as functionals
on the spaces $\Phi_\pm$: the Lippmann-Schwinger kets $|E\mp
i\epsilon, jj_3\eta\r>$ can be analytically extended into the complex
semi-plane ${\mathbb{C}}_\mp$ (this means the bra
$\<^+\overline{E+i\epsilon},jj_3\eta|$ and the ket $|E-i\epsilon,
jj_3\eta\r>$ as well as the integrand in the scalar product
$(\psi^-,\phi^+)$ can be analytically extended into the lower
semiplane ${\mathbb{C}}_-$ of the second sheet of the $S$-matrix
$S_j(E)$ except at singularities). The Gamow vectors are the
evaluation of the analytically extended kets $|zjj_3\eta^-\r>$ in the
lower half plane at the position $z_R=E_R-i\frac{\Gamma}{2}$ of the
first order $S$-matrix pole.  (Gamow-Jordan vectors belong to the
higher order poles \cite{gamow-jordan}.)  Then, from \eqref{3.9}, the
time evolution of the Gamow vectors is given by
\begin{eqnarray}
  e^{-iH_+^\times
    t}|z_Rjj_3\eta^-\r>&=&e^{-iz_Rt}|z_Rjj_3\eta^-\r>\nonumber\\
  &=&e^{-iE_Rt}e^{-\frac{\Gamma}{2}t}|z_Rjj_3\eta^-\r>\ \text{for
    $t\geq0$ only.}\label{3.10}
\end{eqnarray}
This means there is an association between the the resonance pole of
the $j$-th partial scattering amplitude $a_j(E)$ and the Gamow
vectors:
\begin{equation}
  \left.
    \begin{matrix}
      \text{Resonance pole at $z_R=E_R-i\frac{\Gamma}{2}$}\\
      \text{described by}\ a_j(E)=\frac{r}{E-z_R}
    \end{matrix}
  \right\}\quad
  \Longleftrightarrow\quad
  \left\{
    \begin{matrix}
      \text{Space of states of Gamow}\\
      \text{vectors spanned by}\ |z_Rjj_3\eta^-\r>
    \end{matrix}\right.
  \label{3.10.1}
\end{equation}
The resonance is defined by a pole of the $S$-matrix element of
angular momentum $j$ at the complex energy $z_R=E_R-i\frac{\Gamma}{2}$
and is measured as a Lorentzian (Breit-Wigner) bump with maximum at
$E_R$ and full width at half-maximum~$\Gamma$:
\begin{equation}
  \left|a_j(E)\right|^2=
  \frac{|r|}{(E-E_R)^2+\left(\frac{\Gamma}{2}\right)^2}
  \label{3.10.2}
\end{equation}
To this resonance corresponds a ket which is defined by the Cauchy
integral around the $S$-matrix pole $z_R$
\begin{equation}
  |z_Rjj_3\eta^-\r>=\frac{1}{2\pi i}\oint dz\frac{|zjj_3\eta^-\r>}{z-z_R}
  =\frac{i}{2\pi}\int_{-\infty_{II}}^\infty
  dE\frac{|Ejj_3\eta^-\r>}{E-z_R}\label{3.10.3}
\end{equation}
The second equality of \eqref{3.10.3} is the Titchmarsh theorem for
Hardy functions (written here for functionals). This equality and the
association \eqref{3.10.1} between Breit-Wigner resonance amplitude
and Gamow state therefore require the new axiom (\ref{3.7}$\pm$).
\eqref{3.10.3} expresses the new ket $|z_Rjj_3\eta^-\r>$ by a Dirac
basis vector expansion as in (A3), except that the continuous
summation extends over all real energy values $-\infty_{II}<E<\infty$,
where $-\infty_{II}$ means that for the ``unphysical'' values $E<0$,
the energy $E$ is on the second Riemann sheet. We call the ket
\eqref{3.10.3} with the energy wave function given by the Breit-Wigner
amplitude \eqref{3.10.1} a Gamow vector because one can prove (again,
using axiom (\ref{3.7}$\pm$) that it fulfills \eqref{3.10}.  This
Gamow vector \eqref{3.10.3} provides a state vector description to the
Breit-Wigner resonance \eqref{3.10.2}. The semigroup time evolution
\eqref{3.10} of this state vector shows that this state is
exponentially decaying with a lifetime $\tau=\frac{1}{\Gamma}$, where
$\Gamma=-2\Im(z_R)$.

Unstable particles that are characterized by their lifetime are called
decaying states, and they are conceptually and experimentally
different from resonances, which are characterized by the resonance
energy and width. From \eqref{3.10} and the fact that $z_R$ is the
$S$-matrix pole, we see that the Gamow vector provides a unified
description of decaying states and resonances, which can now be
collectively called quasistable states. They elevate the heuristic
lifetime-width relation $\tau=\frac{1}{\Gamma}$ to an exact and
universal identity between two quantities that are observationally and
mathematically different.

The time evolution equations \eqref{3.2}, \eqref{3.8}, \eqref{3.9} and
\eqref{3.10} imply a particular finite value $t=0$ at which time
begins. What is the physical meaning of this initial moment of time?
To answer the question, notice that under the axiom \eqref{3.7}, the
Born probabilities $\Pc_{\phi^+}(\psi^-(t))$ are defined, due to
\eqref{3.7.5-}, only for $t\geq0$:
\begin{equation}
  \Pc_{\phi^+}(\psi^-(t))=\left|\<\psi^-(t)|\phi^+\r>\right|^2=
  \left|\<\psi^-|\phi^+(t)\r>\right|^2\ \text{for $t\geq0$
    only.}\label{3.11}
\end{equation}
For a resonance or decaying state represented by a Gamow vector
$|z_R^-\r>$, we have, using \eqref{3.10},
\begin{equation}
  \Pc_{|z_R\r>}(\psi^-(t))=\left|\<\psi^-(t)|z_R^-\r>\right|^2=
  \left|\<\psi^-|z_R^-(t)\r>\right|^2=e^{-\Gamma
    t}\left|\<\psi^-|z^-_R\r>\right|^2\ \text{for $t\geq0$ only.}\label{3.12}
\end{equation}
Equations \eqref{3.11} and \eqref{3.12} tell us that a time
independent observable $\psi^-$ can be measured in a time dependent
state $\phi^+(t)$ only after a particular instant $t=0$.  (or,
equivalently, the time dependent observable $\psi^-(t)$ can be
measured in a time independent state $\phi^+$ only {\em after} the
same instant $t=0$). In the case of the quasistable state of
\eqref{3.12}, the time $t=0$ is interpreted as the time at which the
state $|z_R^-(t)\r>$ has been prepared, i.e., the quasistable particle
is produced or formed. The observable $|\psi^-(t)\r>\<\psi^-(t)|$
representing the decay products can be detected only after this time,
$t\geq0$. From this point of view, the semigroup condition $t\geq0$
expresses a simple causality condition: The observable $\psi^-$ can be
measured only at times $t$ larger than the time $t=0$ at which the
state is prepared.

Such a particular moment {\em cannot} be singled out if we instead use
the unitary group evolution of the Hilbert space, for which the
probabilities \eqref{3.11} are necessarily defined for all
$-\infty<t<\infty$. It is well known that there are serious problems
with accommodating causality into the conventional formalism of
quantum mechanics~\cite{fermi}. Therefore, the causal time evolution
that follows from the new Hardy space axiom is welcome. But it also
poses a new question: what is the meaning of the semigroup time $t=0$
and how can we observe it? This will be discussed in the following
section.

\section{Observing the Semigroup Time of Causal
Evolution}\label{sec4}

The causal quantum mechanical semigroup (\ref{3.2}$\pm$) introduces a
new concept, the semigroup time $t=t_0$. In the mathematical
description, we call this $t_0=0$, but physically $t_0$ could be any
finite time $(\not=-\infty)$. This concept of a beginning of time is
foreign to the conventional mathematical theory of quantum physics
based on the Hilbert space axiom (A2) (or its slightly strengthened
version \eqref{2.6}), in consequence of which follow the time
evolution equations (\ref{2.8}$\pm$) with $-\infty<t<+\infty$.
Nevertheless, a beginning of time $t_0$ has been mentioned before by
Gell-Mann and Hartle in their quantum theory of the universe
\cite{gell-mann}, where $t_0$ was chosen as the big bang time and
where the restriction of the unitary group evolution \eqref{2.8-} to
$(t-t_{\rm big bang})\geq0$ was introduced by fiat, in contradiction
to the prediction~(\ref{2.8}$\pm$) of the Hilbert space axiom~(A2). In our
theory presented in this paper, the time asymmetry (\ref{3.2}$\pm$) is
a consequence of our Hardy space axiom (\ref{3.7}$\pm$) which was
demanded by the heuristic ($\mp i\epsilon$) in scattering theory (and
also in the propagator of field theory).

We now want to answer the questions: what is the meaning of this
beginning of time $t_0$ for quantum systems in experiments in the
laboratory, and why have we not been more aware of its existence
before?

\begin{figure}[H]
  \centering \includegraphics[width=0.7\textwidth]{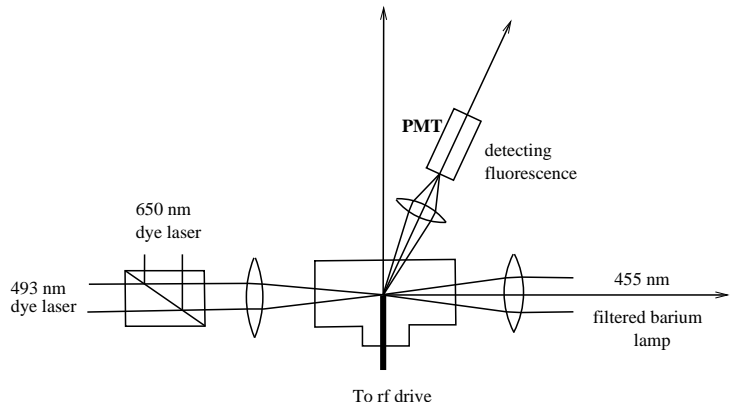}
  \caption{Schematics of the experimental setup used
    in~\cite{dehmelt,sauter}}
  \label{fig:1}
\end{figure}

\begin{figure}[H]
  \centering \includegraphics[width=0.7\textwidth]{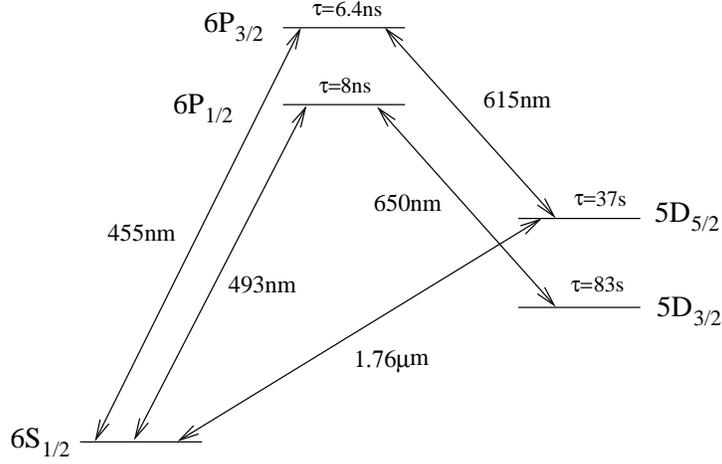}
  \caption{Simplified energy-level scheme of $Ba^{+}$.}
  \label{fig:2}
\end{figure}

In the usual experiments with quantum systems one works with a large
ensemble. For example, the preparation time of an excited state of an
atom or ion corresponds to the many different laboratory clock times
at which each individual atom or ion of the ensemble is created. The
situation is different if one can work with single quantum systems. By
now, there are several experiments that use single, laser-cooled ions
\cite{dehmelt,sauter}. The original experiments used $Ba^+$ in a
Paul-Straubel trap, Fig.~\ref{fig:1}.  This is one of the simplest
cases that nature provides with the most suitable arrangements for
resonance energy levels and lifetimes, as depicted in
Fig.~\ref{fig:2}.

\begin{figure}[H]
  \centering \includegraphics[width=\textwidth]{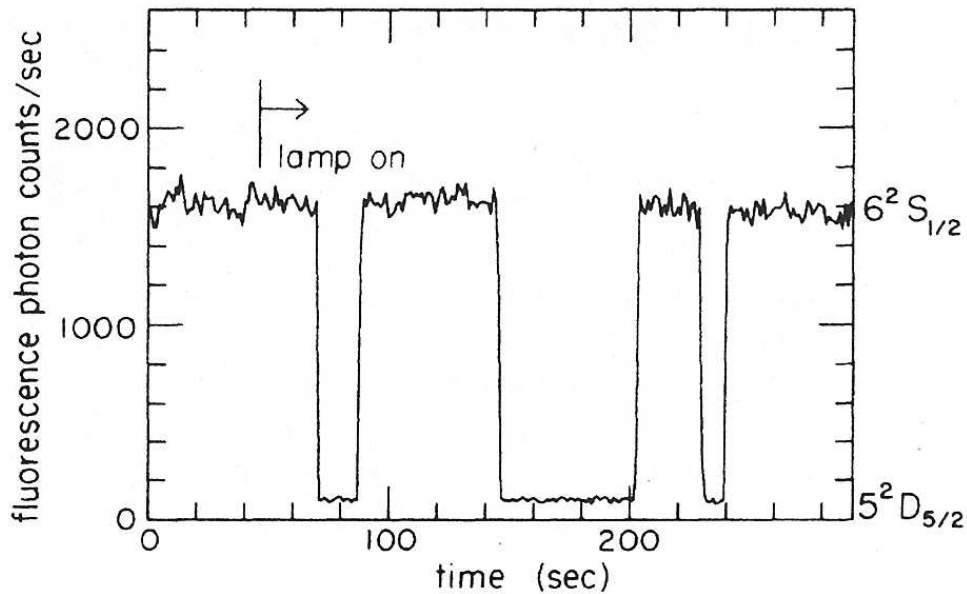}
  \caption{Amplification of single quantum jumps by the fluorescence
    of $S_{1/2}\longleftrightarrow P_{1/2}$.  The 203 onset times
    $t^{i}_{0}$ (three shown) of dark fluorescence are the preparation
    times of single ``$D_{5/2}$''--quantum systems. The ensemble
    $\{t^{i}_{0}\}$ is the preparation time $t_{0}=0$ of the decaying
    quantum state $|5\,D_{5/2}^{\;\;\;\;\;\;-}\rangle$ which
    represents this ensemble of ``$D_{5/2}$''--quantum systems.
    Usually the quantum mechanical ensemble is thought of as a large
    number of micro-objects at one and the same time. Here, the
    quantum mechanical ensemble is one and the same micro object
    prepared at a large number of times $t_{0}^{i}$, which is the same
    time $t=0$ of the state $|5\,D_{5/2}^{\;\;\;\;\;\;-}\rangle$
    describing the ensemble (this figure is taken from
    Ref.~\cite{dehmelt}.)}
  \label{fig:3}
\end{figure}

In these experiments a single laser-cooled $Ba^+$ ion in a trap
undergoes two laser driven transitions. First, driven by the 493-nm
dye laser (Fig.~\ref{fig:1}), the ion goes from the ground state
$6S_{1/2}$ into the excited state $6P_{1/2}$ from where it almost
instantaneously (8 ns) decays into state $5D_{3/2}$. Second, from
state $5D_{3/2}$ the ion is driven back to state $6P_{1/2}$ by the
650-nm dye laser (Fig.~\ref{fig:1}), from where it decays into the
ground state, emitting 493-nm fluorescence radiation. This
fluorescence radiation is monitored by the photo multiplier tube (PMT)
in Fig.~\ref{fig:1}. Initially, the intensity of the fluorescence
radiation shown in Fig.~\ref{fig:3} is essentially a constant at about
16,000 counts/sec. Then, at the time ``lamp on'', a 455-nm filtered
Barium lamp (Fig.~\ref{fig:1}) is turned on. After this ``lamp-on''
time, the fluorescence radiation changes rapidly at random times from
the initial value of 16,000 counts/sec to the background value of no
fluorescence. The explanation is the following: The Barium lamp
occasionally excites the $Ba^+$ into the state $6P_{3/2}$ from where
it makes a fast transition into the state $5D_{5/2}$. This is a
metastable state described by the Gamow vector $|z_R\ 5{D_{5/2}}^-\r>
\equiv\psi^G$. Since there is only one $Ba^+$ atom, it can either go
through the transition levels
$6S_{1/2}\leftrightarrow6P_{1/2}\leftrightarrow5D_{3/2}$ or be
``shelved'' in the metastable state $5D_{5/2}$. While it is shelved
there cannot be fluorescent radiation $6P_{1/2}\rightarrow6S_{1/2}$,
which results in a dark period.

The experiment \cite{dehmelt} reported 203 dark periods, of which
three are shown in Fig.~\ref{fig:3}. The state vector $\psi^G$
represents the ensemble of these 203 single quantum systems. (The
superscript $^-$ in $\psi^G=|z_R\ 5{D_{5/2}}^-\r>$ indicates that this
is an eigenstate of the total Hamiltonian $H=H_0+H_I$, including the
interaction $H_I$ and thus not an eigenstate of the orbital angular
momentum with $(L=2)=D$.) The state $\psi^G$ evolves in time according
to \eqref{3.10} and decays exponentially in time according to
\eqref{3.12}.)  Fig.~\ref{fig:3} shows that each of the single systems
making up the ensemble described by the state vector $|z_R\
5{D_{5/2}}^-\r>=~\psi^G$ is individually produced by the resonance
production process
\begin{equation}
  \gamma(455\text{-nm})+6S_{1/2}\rightarrow6P_{3/2}
  \rightarrow\gamma(615\text{-nm})+5D_{5/2}
  \label{4.1}
\end{equation}
at particular laboratory times $t_0^1,\ t_0^2,\ t_0^3,\cdots,\
t_0^{203}$. (Of these, $t_0^1,\ t_0^2$ and $t_0^3$ are shown in
Fig.~\ref{fig:3} as the onset time of the first three dark periods.)
These excited ions in $5D_{5/2}$ then decay according to
\begin{equation}
  5D_{5/2}\rightarrow6S_{1/2}+\gamma(1.76\text{-$\mu$m})\label{4.2}
\end{equation}
at times $t_1^1,\ t_1^2,\ t_1^3,\ \cdots,\ t_1^{203}$, the instances
at which the fluorescence returns to its pre-``lamp-on'' levels. The
duration of the dark period $\Delta t^{i}=t_1^i-t_0^i,\
i=1,2,3,\cdots,203$, is the time which the $i$-th individual quantum
system $5D_{5/2}$ ``lives''. That is, at every onset time $t_0^i$ of
the $i^{\rm th}$ dark period, the accuracy of which is determined by
the short production time of \eqref{4.1}, an individual $5D_{5/2}$ is
``created''. It ``lives'' for the duration $\Delta t^i=t_1^i-t_0^i$
and decays at $t_1^i$, the end of the $i^{\rm th}$ dark period.

This is a rather remarkable observation because it means that the
excited $Ba^+$ in the quasistable $5D_{5/2}$-level lives for a precise
time $\Delta t^i$. However, these times $\Delta t^i$ are {\em not
  reproducible} quantities, as seen from the different duration
lengths of the dark fluorescence periods.

The reproducible quantity is the ensemble average of the time
intervals $\Delta t^i$, the lifetime of the state $5D_{5/2}$:
\begin{equation}
  \tau^{\rm exp}=\sum_i\Delta t^i\frac{N_D(t:\ \Delta t^i>t)}{N_D}.
\label{4.3}
\end{equation}
Here, $N_D(t:\ \Delta t^i>t)$ is the number of dark periods of
duration $\Delta t^i>t$ and $N_D$ is the total number of dark periods
(203 for this experiment). In the Gamow vector description of the
quasistable state $5D_{5/2}$, a theoretical prediction of the quantity
$\tau^{\rm exp}$ can be made in terms of the resonance width, as shown
below. The individual times $\Delta t^i$ are {\em not predictable}
quantities in quantum mechanics.

Let us now turn to the description of the state $5D_{5/2}$ by the
Gamow state $\psi^G$ and the problem of the physical meaning of the
beginning semigroup time $t_0$. As discussed above, the ensemble state
$5D_{5/2}$ consists of a large number of individual quantum physical
systems, each created at a different laboratory time $t_0^i$. These
times depend on the preparation conditions such as the intensity of
the barium lamp (in the present experiment, it is chosen such that a
transition to $P_{3/2}$ takes place once every 10 s). However, as seen
from \eqref{4.3}, the reproducible experimental quantities depend only
on the time intervals $\Delta t^i$, and not on the individual creation
times $t_0^i$ or the decay times $t_1^i$. The time interval $\Delta
t^i=t_1^i-t_0^i$ is clearly invariant under a translation by $t^i$,
i.e., $\Delta t^i=t_1^i-t_0^i=(t_1^i-t^i)-(t_0^i-t^i)$. Now, a time
$t^i$ can be chosen for each laboratory creation time $t_0^i$ such
that
\begin{equation}
  t_0^i-t^i=t_0\label{4.4}
\end{equation}
where the time $t_0$ is independent of the index $i$. The particular
choice $t_0=0$ (i.e., $t^i=t_0^i$) corresponds to the beginning
semigroup evolution time of the Gamow state $\psi^G$.

What \eqref{4.4} shows, above all, is that the individual
micro-physical systems that make up an ensemble described by a quantum
mechanical state can be prepared at different times (and, for that
matter, different points in space). The time $t_0=0$ of \eqref{4.4}
provides a reference time for the entire ensemble of the creation
times $\left\{t_0^i\right\}$,
\begin{equation}
  \left.\begin{matrix}
      \text{Ensemble of experimental}\\
      \text{ preparation times}
      \{t_0^i\}
    \end{matrix}\right\}
  =
  \left\{\begin{matrix}
      \text{Theoretical semigroup time}\\
      t_0=0\ \text{of the prepared state}
    \end{matrix}\right.
  \label{4.5}
\end{equation}
Thus, the individual systems of the ensemble can be treated as if they
were created at the same laboratory time and the duration that each
micro system ``lives'' can simply be characterized by the time at
which it decays. This feature makes it possible to describe the entire
ensemble by a single Gamow state vector $\psi^{G}$ and the time
evolution of the entire ensemble by a single time variable $t\geq
t_0$. Such a state vector description, in turn, makes it possible to
use the standard probability interpretation also for an ensemble that
consists of a large number of micro systems created at vastly
different laboratory times. For instance, by using \eqref{3.10} for
the Gamow vector $\psi^G(t)=e^{-iHt}|z_Rjj_3\eta^-\r>=e^{-iHt}|z_R
5{D_{5/2}}^-\r>$, the lifetime of the excited state $5D_{5/2}$ can be
computed in analogy to \eqref{3.12} as:
\begin{equation}
  \tau^{\rm theor}=\int_{t_0=0}^\infty dte^{-\Gamma
    t}=\frac{1}{\Gamma}\label{4.6}
\end{equation}
The experimental quantity of \eqref{4.3} is to be compared with this
theoretical quantity.

New in these remarkable experiments of \cite{dehmelt, sauter} is that
the different creation times $t_0^i$ and durations times $\Delta t^i$
for the single quantum systems are precisely and individually measured
as the onset and duration of the dark periods of Fig.~\ref{fig:3}.
These onset times are an experimental demonstration of the semigroup
time $t_0=0$ of time asymmetric quantum theory.

\section{Summary}\label{sec5}

Many of the heuristic notions used in the description of scattering
and decay phenomena, like the incoming and outgoing Lippmann-Schwinger
kets $|E^\pm\r>=|E\pm i\epsilon\r>$ with infinitesimal $\epsilon$,
purely outgoing boundary conditions, time asymmetry and causality are
not well defined in the mathematical frame set by the conventional
(Hilbert space) quantum mechanics. Combining these notions with the
Hilbert space axiom leads to contradictions, like the exponential
catastrophe in which Gamow vectors and unitary time evolution
conflicted \cite{bohm2}, the deviations from the exponential decay
where the exponential time dependence for the experimental counting
rates conflicted with the mathematical properties of Hilbert space
vectors \cite{bohm3}, and the problems with (Einstein) causality where
stability of matter (semi-boundedness of the Hilbert space
Hamiltonian) leads to instant propagation of probabilities
\cite{fermi}. The $\pm i\epsilon$ of the Lippmann-Schwinger kets (or,
of the propagator in relativistic quantum field theory) overcomes many
of these problems.

But the Lippmann-Schwinger kets are mathematically undefined kets;
they are not vectors of the Hilbert space and they cannot be defined
as Schwartz space functionals because of the $\pm i\epsilon$.
Therefore one cannot derive their time evolution (or, in the
relativistic case, their evolution under Poincar\'e transformations).
Nevertheless, one {\em assumes} it to be a unitary time evolution (as
one also had {\em assumed} for the ordinary Dirac kets) with time
extending over $-\infty<t<\infty$. This however is in conflict with
the infinitesimal imaginary part $\pm i\epsilon$ since it would lead
to non-continuous and unbounded (non-unitary) operators for time
evolution (or, in the relativistic case, non-unitary representations
of the Poincar\'e group). Complex extensions of energy (or, in the
relativistic case, the invariant square mass $s=p_\mu p^\mu$) away
from the real axis requires that the energy wave functions be boundary
values of analytic functions in the complex semi-planes, not just
(Lebesgue) square-integrable or smooth functions of real energy.

Using the Lippmann-Schwinger equation as the takeoff point and
attempting to accommodate as many of the heuristic notions of
scattering and decay as possible, we conjectured in this paper the new
hypothesis (\ref{3.7}$\pm$).  It replaces the Hilbert space boundary
conditions (A2) for the solutions of the Schr\"odinger or Heisenberg
equation by the Hardy space boundary conditions (\ref{3.7}$\pm$).
Many of the heuristic notions, such as Gamow's wave functions, that
had been introduced phenomenologically into the description of
scattering and decay phenomena appear also in this new quantum theory,
but now they have a rigorous mathematical foundation. Furthermore, the
new theory leads to important novel conclusions, salient among which
is a basic, quantum mechanical time asymmetry, expressed by the
semigroup evolution of (\ref{3.7.5}$\pm$). This overcomes the
causality problem and leads to exponential decay for certain kets with
complex energy, the Gamow kets.

Gamow kets have been derived from the resonance poles of the
$S$-matrix using the new axiom (\ref{3.7}$\pm$), Their energy wave
function is a Lorentzian (Breit-Wigner) energy distribution
characterized by its central value $E_R$ and width $\Gamma$, and the
lifetime of its exponential decay is exactly
$\tau=\frac{\hbar}{\Gamma}$. The new axiom (\ref{3.7}$\pm$) thus
provides a unified theory of resonance scattering and exponential
decay.

But the semigroup also introduces a beginning of time for quantum
systems, which is represented by the mathematical semigroup time
$t=0$. Though such a time has been mentioned before as the big bang
time for universes \cite{gell-mann} and its idea is already contained
in the classic paper \cite{feynman}, one has not been much aware of it
in the usual experiments with quantum systems in the laboratory. In
the final section \ref{sec4}, we therefore discussed an experiment
with single laser-cooled $Ba^+$ ions in a trap \cite{dehmelt} where
the beginnings of time for single micro-systems have been observed.

\end{document}